\documentclass[10pt,a4paper,twoside]{article}
\usepackage[a4paper,left=30mm,right=30mm, top=1cm, bottom=2cm]{geometry}

\usepackage[T1]{fontenc}
\usepackage[utf8]{inputenc} 
\usepackage[english]{babel} 
\usepackage{amsmath, amsthm, amssymb}
\usepackage{authblk}
\usepackage{graphicx}
\usepackage{tikz}
\usepackage{caption, subcaption}
\usepackage[nottoc,numbib]{tocbibind}
\usepackage{hyperref}
\usepackage{pdfpages} 

\date{\today}

\addto\extrasngerman{}
\addto\extrasngerman{}
\addto\extrasngerman{}
\addto\extrasngerman{}
\addto\extrasngerman{}

\author{Christian Berghoff}
\affil{Bundesamt für Sicherheit in der Informationstechnik, Bonn, Germany}
\affil{firstname.lastname@bsi.bund.de}
\title{Protecting the integrity of\\ the training procedure of neural networks}
\date{\today}

\begin{document}

\maketitle
\begin{abstract}
Due to significant improvements in performance in recent years, neural networks are currently used for an ever-increasing number of applications. However, neural networks have the drawback that their decisions are not readily interpretable and traceable for a human. This creates several problems, for instance in terms of safety and IT security for high-risk applications, where assuring these properties is crucial. One of the most striking IT security problems aggravated by the opacity of neural networks is the possibility of so-called poisoning attacks during the training phase, where an attacker inserts specially crafted data to manipulate the resulting model. We propose an approach to this problem which allows provably verifying the integrity of the training procedure by making use of standard cryptographic mechanisms.\\
\end{abstract}


\section{Introduction}
The lack of transparency and interpretability in neural networks is one of the root causes of their problems from the perspective of safety and IT security \cite{paper}. Starting from an initial state, a neural network is trained on training data using a certain algorithm. This gives rise to a new state of the neural network usually better fitted to solve the problem at hand. However, it is in general not possible to derive any reliable statements from this new state on what happened during training. Hence, it is extremely difficult to detect malicious data introduced by an attacker into the training data set in order to manipulate the resulting model in a so-called poisoning attack \cite{Chen:2017}. Although tests may be conducted after training using the new state of the network, due to the typically large dimension of the input space it is highly unlikely that they will reveal the effects of specially crafted poisoned data.\\

This document presents a different approach to the problem. Essentially, it allows verifying the correctness and integrity of the training procedure for a neural network afterwards. Hash functions, which are a standard cryptographic tool, are used to make sure an attacker cannot tamper with training data and claim that other data were used for training than those that were actually used. It is important to note that the ability to verify the training procedure is a large asset, but is not sufficient on its own to reliably detect and counter attacks during the training phase. Indeed, the approach only guarantees that the data provided were actually used for training the neural network. However, it does not make any statement about the integrity of the data themselves and the absence of manipulations, which still need to be checked. This can be done either completely by hand, which is typically unrealistic, by an automatic procedure or by using some technical pre-processing and manual inspection of a much smaller number of data items. Several methods have been proposed to detect poisoning attacks \cite{Chen2018b, Tran2018, Wang2019}. They essentially rely on clustering algorithms, which may take into account not only the data themselves but also the behaviour of the neural network when processing these data. Although some of these methods work quite well under certain circumstances, they cannot yet be used as a reliable tool for detection, especially when facing more sophisticated poisoning attacks, e.g.\ \cite{Saha2019, Turner2019}.
\\

The document starts with some definitions concerning neural networks and briefly reviews hash trees, the main building block of the proposed approach. It then presents a solution which allows verifying the correctness and integrity of training for a neural network in \autoref{sec:verification}. Two cases are distinguished. The straightforward solution from \autoref{sec:cv} targets the whole training procedure, whereas a modified version, presented in \autoref{sec:pv}, allows selectively verifying some parts of the training procedure. The raison d'être of the modification is that the standard solution requires completely redoing the training procedure for the purpose of verification, which may come with very high costs. The modified solution takes samples, and can thus balance the costs by choosing the desired proportion of the training procedure to be checked. However, this partial verification procedure is susceptible to certain attacks, which are analysed in \autoref{sec:sec_ana}. The solutions outlined in \autoref{sec:verification} are extended in greater technical detail in \autoref{sec:app}.

\section{Definitions}\label{sec:def}
In this document, we consider a neural network to be a function $f: X \rightarrow Y$ mapping data from the input space $X$ to the output space $Y$, which is defined by its general set-up (including the type of network and the layers used) and the set of weights $(w_i)_{i=1}^n \in \mathbb{R}^n$ connecting its neurons.\\

In the training phase of the neural network, the initial weights $w_i \in \mathbb{R}, i=1, \ldots, n$, are updated using (batches of) data from the training set $D=\{D_i: i=1, \ldots, d\}$, while the general set-up is fixed. One training step may thus be formalised as a function
\[
T: \mathbb{R}^n \times D^b \rightarrow \mathbb{R}^n,\quad ((w_i)_{i=1}^n, (D_{j_1}, \dots, D_{j_b})) \mapsto (v_i)_{i=1}^n,
\]
which induces a function
\[
T: Map(X, Y) \times D^b \rightarrow Map(X,Y),\quad (f, (D_{j_1}, \dots, D_{j_b})) \mapsto g,
\]
where the set-up of $g$ is the same as that of $f$ and the weights of $g$ are $(v_i)_{i=1}^n$ and $b \in \mathbb{N}$ is the batch size, which is usually fixed. We use the shorthand notation $\bar{D}_j=\bar{D}_{j_1, \dots, j_b}=(D_{j_1}, \dots, D_{j_b})$ for a batch of data. The function $T$ may implicitly depend on certain values, e.g. pseudo-random seeds, from which pseudo-random numbers are derived, or hyperparameters, cf.\ \autoref{sec:data_details}. Furthermore, we define the map $S: \{1, \ldots, d\} \rightarrow D, i \mapsto D_i$ for numbering the training data.\\
A hash tree \cite{Merkle89} is a standard cryptographic data structure. In a hash tree, each leaf contains some data (or its hash value), and each node contains the hash value of some combination (typically, the concatenation) of the contents of its children. Using hash trees, one can efficiently verify the integrity of large amounts of data. In order to verify the presence of a leaf $a_0$ in a hash tree, one needs to compute all intermediate hashes until one reaches the root hash $h_{root}$ of the tree. This requires computing a logarithmic (in the number of leaves) number of hashes if the number of children of a node is constant (e.g.\ two for binary hash trees). More precisely, for a binary hash tree, given a leaf $a_0$, and the root hash $h_{root}$, one needs to compute the hashes $a_1:=H(a_0 | b_0), a_2:=H(a_1 | b_1), \ldots, a_{\ell+1}:=H(a_\ell | b_\ell)$ and check that $a_{\ell+1}=h_{root}$ holds, where $H$ is the hash function used, the tree has depth $\ell+2$, $b_0, \ldots, b_\ell$ is the sibling (in this example, the right sibling) of $a_0, \ldots, a_\ell$, respectively, and concatenation of $a$ and $b$ is denoted by $a | b$. Besides $a_0$ and $h_{root}$, the values $b_i, i=0, \ldots, \ell$, are thus necessary to verify that $a_0$ is contained in the tree with root hash $h_{root}$. Generalising the verification procedure to non-binary hash trees is straightforward. At each level, it requires the values of all siblings of the node through which the hash chain to the root passes.\\
We assume that the hash tree uses a cryptographically secure hash function $H$. In addition, we make the assumption that the root hash $h_{root}$ is protected using a secure digital signature algorithm and that the signature itself is properly authenticated (e.g. via a PKI), which prevents tampering. Recommendations for these functions may for instance be found in \cite{TR_krypto}.

\section{Verification procedure}\label{sec:verification}
This section outlines the procedures for a complete or partial verification of the training procedure. Both solutions use hash trees. In principle, the complete verification, which needs to use all the data and completely recalculate the hash tree, could also use other hash-based data structures, like hash lists. However, for the sake of consistency we present the solution based on hash trees, since the partial verification does rely on properties of hash trees. In each solution, the proving party needs to store data specifying the training procedure, merge it into a hash tree and sign its root hash. Upon request by the verifying party, the proving party provides some information from the hash tree as well as the relevant information from the training procedure. This allows the verifying party to check the training procedure was conducted as stated by the proving party.

\subsection{Complete verification}\label{sec:cv}

We denote our function with the initial set of weights $f_0$. This function is now trained by repeated application of the function $T$, where the second argument $\bar{D}_j \in D^b$ may differ between iterations, but $b$ is usually fixed as stated in \autoref{sec:def}. This gives rise to a chain of transformations
\[
f_0 \mapsto T(f_0, \bar{D}_{i_{0, 1}, \dots, i_{0, b}})=:f_1 \mapsto \ldots \mapsto T(f_{k-1}, \bar{D}_{i_{k-1, 1}, \dots, i_{k-1, b}})=:f_{k}
\]
with $i_{0, 1}, \ldots, i_{k-1, b} \in \{1, \ldots, d\}$. Using the map $S$ and the function $T$, knowledge of $f_0$ and the indices $\{i_{0, 1}, \ldots, i_{k-1, b}\}$ completely determines $f_k$.\\
This observation gives rise to the following idea. We use a hash tree $h$ whose leaves contain the following information (for more details, see \autoref{sec:data_details}):
\begin{enumerate}
\item Meta data
\item Information determining the set-up of $f_0$ 
\item Information determining the function $T$
\item Information determining the map $S$
\item The indices $i_{0, 1}, \ldots, i_{k-1, b}$ used for training $f_0$
\item The initial weights $(w_i)_{i=1}^n$ of $f_0$
\end{enumerate}
The data $D$ themselves need not be directly included in the hash tree, but rather their hash values are stored in category 4.\ (see \autoref{sec:data_details} for details). We assume that these data are in any case stored by the proving party as a backup or for future use, whether or not the information for subsequent verification is generated during the training procedure.
\\

Then given a neural network $f_k$ with weights $(v_i)_{i=1}^n$ and the digitally signed root hash $h_{root}$ of $h$, the verifying party can check that $f_k$ was derived from $f_0$ using the training procedure as specified by $h$. Since the result of training is deterministic given all the information contained in the leaves of $h$, it is not possible to provide false information on the data used for training or on the method applied.\\

More precisely, the verifying party will check the following conditions in the stated order:
\begin{enumerate}
\item The digital signature of $h_{root}$ is authentic and correct.
\item Hashing the information mentioned above in the right way gives $h_{root}$.
\item The training data $D$ hashes to the values stored in 4.
\item Applying the training procedure as specified by $h$ to $f_0$ with initial weights $(w_i)_{i=1}^d$ yields $f_k$.
\end{enumerate}

\subsection{Partial verification}\label{sec:pv}
While the straightforward approach from \autoref{sec:cv} can be used for the purpose of verification, it would require a lot of computational resources. The resources for the first three steps can be assumed to be negligible in comparison, but the fourth step requires the verifying party to repeat all the computations necessary to reach the final state $f_k$ starting from $f_0$. One can assume that in many cases this is not acceptable or even infeasible for the verifying party. On the one hand, the computing power required may be prohibitive for large neural networks, even if we assume that the final training, which results in a neural network meeting the developer's goals, accounts for only a small fraction of the total computational effort expended in development. On the other hand, the proving party may not want to disclose all data used for training the network, for instance for protecting its intellectual property or due to data protection requirements. In such a case, one can modify the procedure in a way which allows proving the correctness and integrity of any batch of multiple intermediate training steps. Proving these properties for all intermediate training steps would amount to proving them for the complete training procedure. The batches to be checked can later be chosen by the verifying party. The proving party then needs to provide the data necessary for checking, and the verifying party can use the hash tree $h$ and these data to check the respective batches. Depending on the amount of training steps to be checked, only a small subset of the training data may need to be disclosed, thus largely protecting the proving party’s intellectual property, since hash values from intermediate levels of $h$ do not leak any information on these data.\\

More precisely, we can use additional checkpoints $f_{i_0}, \ldots, f_{i_m}$ between the initial state $f_0$ and the final one $f_k$, where $i_0=0$, $i_m=k$ and $i_{j+1}>i_{j}$ for all $j\in \{0, \ldots, m-1\}$. The number $m$ is a parameter. The concrete value that should be assigned to it depends on the computational effort required for the transition between two checkpoints and the space required for storing a checkpoint.\\
A checkpoint is defined by the tuple $(i_j, f_{i_j}, I_{i_j})$, where $i_j$ is the number of times $T$ was applied to $f_0$ to arrive at $f_{i_j}$ (in other words, the number of training steps), and the neural network $f_{i_j}$ is defined by its weights $(w_{k}^{(i_j)})_{k=1}^n$. $I_{i_j}$ may be empty or, if applicable, hold additional required information. For instance, $I_{i_j}$ may contain the state of the pseudo-random number generator at this point, if this information cannot be straightforwardly derived from the initial pseudo-random seeds (stored in 3., cf.\ \autoref{sec:data_details}) and the value $i_j$ itself.\\
Then given $f_{i_j}$ and its weights for some $i_j \in \{0, \ldots, k\}$, $f_{i_{j+1}}$ and its weights, and the training data used for the transition from $f_{i_j}$ to $f_{i_{j+1}}$, the verifying party can check the correctness and integrity of the training steps for the transition between the checkpoints $f_{i_j}$ and $f_{i_{j+1}}$. This is done by recomputing these steps and checking that by hashing the respective information one ultimately arrives at the root hash $h_{root}$ of the hash tree $h$. As before, a digital signature of $h_{root}$ must also be provided.\\

In this case, the hash tree $h$ includes the following information:
\begin{enumerate}
\item Meta data
\item Information determining the set-up of $f_0$
\item Information determining the function $T$
\item Information determining the map $S$
\item For each $i_j$ with $j \in \{0, \ldots, m\}$:
\begin{enumerate}
\item The value $i_j$ itself, i.e.\ the number of training steps since the start of training to arrive at $f_{i_j}$
\item The weights of $f_{i_j}$
\item The $(i_{j+1}-i_j)b$ indices of data used in the training steps between $f_{i_j}$ and $f_{i_{j+1}}$ (which we define as the empty set for $j=m$)
\item If applicable, additional information from $I_{i_j}$\\
\end{enumerate}
\end{enumerate}

The verifying party will check the following conditions in the stated order:
\begin{enumerate}
\item The digital signature of $h_{root}$ is authentic and correct.
\item The information from 1.--4.\ is contained in the hash tree $h$.
\item For each transition from $f_{i_j}$ to $f_{i_{j+1}}$ to be verified:
\begin{enumerate}
\item The respective information from 5.\ is contained in the hash tree $h$.
\item The training data $D$ used in the training steps between $f_{i_j}$ and $f_{i_{j+1}}$ according to 5.(c) hashes to the values stored in 4.
\item Using the training set items as specified by the respective indices and applying the training procedure as specified in 1.--4.\ to $f_{i_j}$ yields $f_{i_{j+1}}$.\\
\end{enumerate}
\end{enumerate}

Whenever the presence of information in the hash tree $h$ is checked, this is done using the properties of $h$ as discussed in \autoref{sec:def}. The proving party needs to furnish all information from intermediate levels of the hash tree which is necessary for these calculations.\\

When choosing the number $v$ of transitions between checkpoints to be verified, there is a trade-off between efficiency and the integrity guarantees attained. On the one hand, decreasing this number reduces the computing time required to repeat the calculations. On the other hand, when checking less transitions, one only checks the correctness of a smaller portion of the training procedure, and the probability of discovering integrity violations diminishes. In any case, the concrete transitions to be checked must not be known beforehand, since using a pre-defined set of transitions would allow an adversary to hide malicious changes without any risk of being exposed.

\section{Security analysis}\label{sec:sec_ana}
In the case of partial verification, an attacker can provide a certain amount of false data and has some chance that he will not be exposed. Since only a certain amount of checkpoint transitions are verified, if the attacker provides false data for a small number of transitions, his risk of exposure is quite low. Assume there are $m$ checkpoint transitions in total, $v$ of them are verified (using random sampling) and the attacker manipulates the data for $a$ transitions, then the probability $p$ that this will go unnoticed is about
\[p=\prod_{i=0}^{v-1}\left(1-\frac{a}{m-i}\right)\approx\left(1-\frac{a}{m}\right)^{v}=\left(1-\frac{a}{m}\right)^{m\cdot \frac{v}{m}}\approx \exp\left(-a \cdot \frac{v}{m}\right).\]

For instance, if the verifying party chooses to check $v=\sqrt{m}$ transitions and the attacker has manipulated data for $a=\frac{v}{10}$ transitions, we get $p\approx \exp(-0.1)\approx 0.9048$. We note that the precision of the first approximation is quite good, unless $v$ gets close to $m$ (i.e.\ a significant part of the transitions are checked); concerning the second approximation, convergence to the exponential function is quite fast. For example, taking $m=2500$ and again using $v=\sqrt{m}$ and $a=\frac{v}{10}$, the exact formula for $p$ evaluates to $p\approx 0.9038$.\\
The attacker may achieve even higher values for $p$, but whether this is still feasible (since $a\geq 1$ necessarily needs to hold) depends on the values of $m$ and $v$, which he cannot directly influence. In addition, a very small value for $a$ means the attacker can only use a small amount of poisoned samples for training, which might severely degrade the performance of a poisoning attack, rendering it ineffective.\\

However, in case an attacker tampers with the verification data, there is no need for him to only provide false information on which data were used for training during a particular transition (category 5.(c) in \autoref{sec:pv}). He might as well lie about the number of training steps taken during this transition (category 5.(a)). In this way, a much more powerful attack may become feasible, since the attacker can generate the verification data for a particular transition and have it contain an average, inconspicuous number of training steps, whereas in fact that transition included many more steps and introduced a massive amount of poisoned data. Essentially, it is not possible to guarantee the correctness of the number of steps asserted for a transition without performing the verification procedure for this transition.\\

One approach to mitigate this problem would be to use a more sophisticated algorithm for sampling the transitions to be checked. For instance, one might choose those transitions with a higher probability whose initial and final weights and/or performance differ much more than is the case on average. Intuitively, this should increase the probability of finding transitions using an unusually large number of training steps. However, the attacker might additionally tamper with the initial or final weights of transitions adjacent to the transition he originally targeted in order to level out the changes and defeat the verifier's heuristic. For instance, when targeting transition $i$ and restricting manipulations to $a$ transitions as above, he might also change the information from category 5.(b) for transitions $i+1, \ldots, i+a-1$. If this successfully levels out any information the verifier might use, the probability of the attack going unnoticed is again $p$ as computed above.

\section{Conclusion and outlook}
The integrity of the training procedure of neural networks can be protected using well-known cryptographic mechanisms, which also allow another party to verify the integrity afterwards. This document has outlined a proposal on how to adapt the cryptographic mechanisms to the setting in question. While integrity can only be guaranteed with certainty by completely repeating the training procedure, the partial verification procedure as specified above can give a verifier a certain amount of confidence about the integrity and expose an attacker manipulating data to the risk of being detected. There is a trade-off between the level of confidence, based upon the probability of manipulations going unnoticed, and the computational effort and storage space required for verification (see \autoref{sec:storage}), which can be tuned using several parameters (the number of checkpoints to be stored and of transitions to be verified).\\

The more sophisticated attack scenarios on the partial verification procedure presented in \autoref{sec:sec_ana}, which are based on including false meta data about the number of training steps between two checkpoints, and possible mitigations could be further explored both analytically and empirically to derive more accurate estimates on the probability of successful attacks.\\

It is important to note that protecting and verifying the integrity of the training procedure does not in itself prevent poisoning attacks, which introduce specially crafted malicious training data, but is just one building block for solving this problem. Rather, the absence of malicious training data must additionally be confirmed using methods for poisoning detection.\\

This document focused on protecting the integrity of the training procedure as the key stage of the life cycle of neural networks, but the underlying ideas lend themselves to an easy generalisation to other stages of this life cycle. For instance, protecting the integrity of transmitted sensor data, their curation and their pre-processing all the way to the training data set can effectively prevent the addition of poisoned samples, if properly implemented.\\

The security of the proposed solution is based on the security of the cryptographic mechanisms used. These should hence be chosen and implemented with care. In particular, digital signatures \textbf{must} be used for sealing the hash tree against tampering and they \textbf{must} be properly authenticated.

\section*{Acknowledgements}
The author would like to thank Ute Gebhardt and Matthias Neu for carefully proofreading earlier versions of this document and providing valuable suggestions for improvement.

\footnotesize
\bibliography{lit}
\clearpage

\normalsize
\appendix
\section{Appendix: Technical details}\label{sec:app}

\subsection{List of data included in hash tree}\label{sec:data_details}
This section makes a proposal on which data to include in the different categories as sketched in \autoref{sec:cv} and \autoref{sec:pv}. The objective is to include all data which are necessary for deterministically reproducing the training procedure or parts thereof. If applicable, the lists may be extended or redundant information may be removed in concrete implementations. We advocate establishing consensus on a standardised list of data and on the structure of the hash tree (see \autoref{sec:hash_structure}) in order to make the approach interoperable between different parties with no or minimal modifications.
\subsubsection*{Meta data}
The meta data contain all information necessary for parsing and interpreting the other pieces of information. This includes the concrete values of the constants $k$ (the total number of training steps) and $m$ (describing the number of checkpoints) as well as the number of values included in the different categories of data and the way they are structured.

\subsubsection*{Information determining the set-up of \texorpdfstring{$f_0$}{f0}}
This category includes the following data:
\begin{enumerate}
\item General architecture of the neural network
\item Number of layers
\item Number of neurons in each layer
\item Ordering of weights (i.e.\ meta data on the order in which the weights are stored in the hash tree)
\end{enumerate}

\subsubsection*{Information determining the function \texorpdfstring{$T$}{T}}
This category includes the following data (readers unfamiliar with the general terminology of neural networks may for instance refer to \cite{Goodfellow2016}):
\begin{enumerate}
\item Loss function
\item Optimisation method
\item Pseudo-random seeds
\item Regularisation terms
\item Values of hyperparameters
\end{enumerate}

\subsubsection*{Information determining the map \texorpdfstring{$S$}{S}}
This category includes the following data:
\begin{enumerate}
\item Size $d$ of training data set
\item $(i, H(D_i))_{i=1}^d$, tuples of the number $i$ and the hash value of training set item $D_i$
\end{enumerate}

\subsubsection*{The indices used for training \texorpdfstring{$f_0$}{f0} or between checkpoints \texorpdfstring{$f_{i_j}$}{fij} and \texorpdfstring{$f_{i_{j+1}}$}{fij+1}}
This is straightforward.

\subsubsection*{Weights}
This category of data contains the weights of the initial function $f_0$ and the intermediate checkpoints $f_{i_j}$, respectively. It includes:
\begin{enumerate}
\item Number $n$ of weights
\item The weights $w_k$ or $w_k^{i_j}$, $k=0, \ldots, n$, respectively
\item If applicable, weight-dependent information used in the optimisation method (e.g.\ momentum method \cite{momentum}, Adam \cite{Adam})
\end{enumerate}

\subsection{Structuring the hash tree}\label{sec:hash_structure}
The concrete structure of the hash tree $h$ affects the storage space and the amount of computations which are required for performing a verification. Binary hash trees would be the straightforward and standard solution. They offer the great advantage that for verifying the presence of a leaf they require at most one additional value at each level (if applicable, the sibling of the respective node) in order to compute and check the hash chain to the root hash. However, in our application the leaves whose presence is checked are not independent from each other and we often need to check whole batches of leaves at the same time anyway. Furthermore, binary trees require more intermediate levels than trees with more siblings. In particular, more intermediate values need to be stored. For instance, using the formulae for the geometric series it is easy to see that when using a hash tree where every node has four children instead of two, one can reduce the amount of intermediate nodes by up to a factor of three. Including the leaves, this leads to saving about one third of the storage space.\\

Due to this observation, we propose the following structure for the hash tree, which as a side effect tries to give some logical meaning to (some) intermediate nodes and thus make the scheme more easily comprehensible.\\

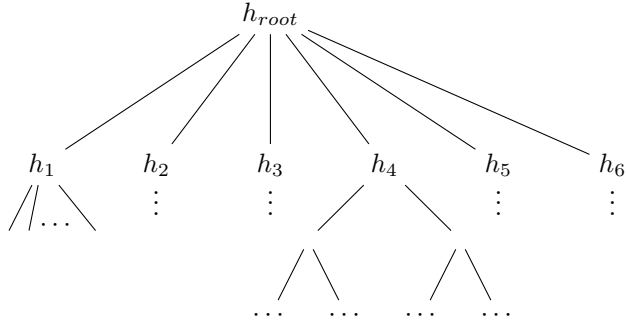
\begin{figure}
\centering
\begin{tikzpicture}
\node (A) at (3,0) {$h_{root}$};
\node (B) at (0,-2) {$h_1$};
\node (B1) at (-0.5, -3) {};
\node (B2) at (-0.2, -3) {};
\node (B3) at (0.2, -2.8) {$\cdots$};
\node (B4) at (0.8, -3) {};
\node (C) at (1.5, -2) {$h_2$};
\node (D) at (3, -2) {$h_3$};
\node (E) at (4.5, -2) {$h_4$};
\node (F) at (6, -2) {$h_5$};
\node (G) at (7.5, -2) {$h_6$};
\node (E1) at (3.5, -3) {};
\node (E2) at (5.5, -3) {};
\node (E11) at (3, -4) {$\cdots$};
\node (E12) at (4, -4) {$\cdots$};
\node (E21) at (5, -4) {$\cdots$};
\node (E22) at (6, -4) {$\cdots$};

\node at (1.5, -2.4) {$\vdots$};
\node at (3, -2.4) {$\vdots$};
\node at (6, -2.4) {$\vdots$};
\node at (7.5, -2.4) {$\vdots$};

\draw (A) -- (B);
\draw (A) -- (C);
\draw (A) -- (D);
\draw (A) -- (E);
\draw (A) -- (F);
\draw (A) -- (G);
\draw (B) -- (B1);
\draw (B) -- (B2);
\draw (B) -- (B4);
\draw (E) -- (E1);
\draw (E) -- (E2);
\draw (E1) -- (E11);
\draw (E1) -- (E12);
\draw (E2) -- (E21);
\draw (E2) -- (E22);
\end{tikzpicture}
\caption{Structure of the hash tree for complete verification (schematic representation). The sub-trees under $h_2$, $h_3$, $h_5$ and $h_6$ are not shown for reasons of space.}
\label{fig:hash_tree_cv}
\end{figure}

For complete verification, all the leaves of the tree need to be checked. In principle, one could imagine a tree with just one level, with the root hash being the hash of the concatenation of all the leaves. However, this would both completely obfuscate the semantic structure of the data to be stored and make debugging overly difficult. Instead, we propose to hash together data from the same category (i.e.\ meta data, information on $f_0$, \dots) and finally compute the root hash from the concatenation of the category-wise root hashes (which we denote $h_j$ for category $j$). Data from the same category may be hashed together using concatenation of all items (which we propose for categories 1.--3., which contain relatively few data) or by using hash sub-trees, whether binary or otherwise (which we suggest for the other categories). A sketch of the hash tree structure is depicted in \autoref{fig:hash_tree_cv}.\\

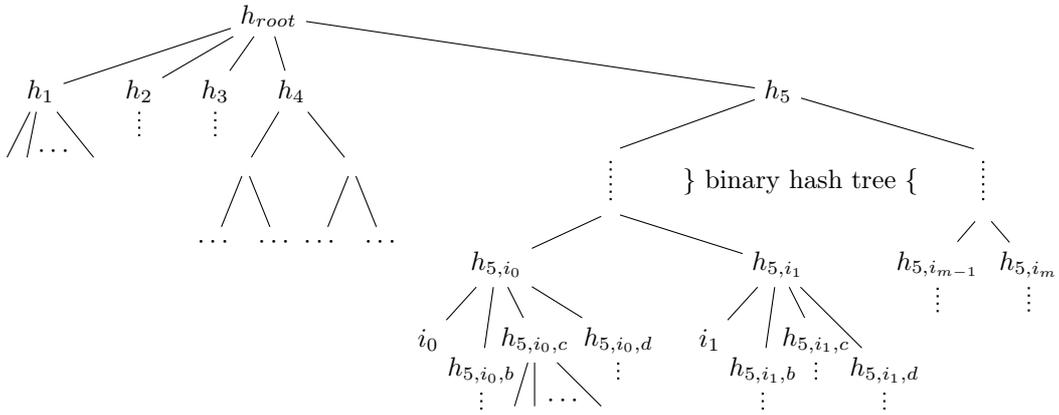
\begin{figure}[ht]
\centering
\begin{tikzpicture}
\node (A) at (3,-1) {$h_{root}$};

\node (B) at (0,-2) {$h_1$};
\node (B1) at (-0.5, -3) {};
\node (B2) at (-0.2, -3) {};
\node (B3) at (0.2, -2.8) {$\cdots$};
\node (B4) at (0.8, -3) {};

\node (C) at (1.3, -2) {$h_2$};
\node (D) at (2.3, -2) {$h_3$};
\node (E) at (3.3, -2) {$h_4$};

\node (F) at (9.7, -2) {$h_5$};
\node (F11) at (7.5, -2.8) {};
\node (F111) at (7.5, -3.6) {};
\node (F12) at (12.4, -2.8) {};

\node (F2) at (6, -4.3) {$h_{5, i_0}$};
\node (F21) at (5.1, -5.3) {$i_0$};
\node (F22) at (5.8, -5.7) {$h_{5, i_0, b}$};
\node (F23) at (6.5, -5.3) {$h_{5, i_0, c}$};
\node (F24) at (7.6, -5.3) {$h_{5, i_0, d}$};

\node (F231) at (6.2, -6.3) {};
\node (F232) at (6.5, -6.3) {};
\node (F233) at (6.9, -6.1) {$\cdots$};
\node (F234) at (7.5, -6.3) {};

\node (F3) at (9.7, -4.3) {$h_{5, i_1}$};
\node (F31) at (8.8, -5.3) {$i_1$};
\node (F32) at (9.5, -5.7) {$h_{5, i_1, b}$};
\node (F33) at (10.2, -5.3) {$h_{5, i_1, c}$};
\node (F34) at (11.1, -5.7) {$h_{5, i_1, d}$};

\node (F121) at (12.4, -3.6) {};
\node (F122) at (11.8, -4.3) {$h_{5, i_{m-1}}$};
\node (F123) at (13, -4.3) {$h_{5, i_m}$};

\node (E1) at (2.7, -3) {};
\node (E2) at (4.1, -3) {};
\node (E11) at (2.3, -4) {$\cdots$};
\node (E12) at (3.1, -4) {$\cdots$};
\node (E21) at (3.7, -4) {$\cdots$};
\node (E22) at (4.5, -4) {$\cdots$};

\draw (A) -- (B);
\draw (A) -- (C);
\draw (A) -- (D);
\draw (A) -- (E);
\draw (A) -- (F);

\draw (B) -- (B1);
\draw (B) -- (B2);
\draw (B) -- (B4);

\draw (E) -- (E1);
\draw (E) -- (E2);
\draw (E1) -- (E11);
\draw (E1) -- (E12);
\draw (E2) -- (E21);
\draw (E2) -- (E22);

\draw[dotted, thick] (F11) -- node[right] {\hspace{.7cm} \} binary hash tree \{} ++ (F111);
\draw (F) -- (F11);
\draw (F) -- (F12);
\draw (F111) -- (F2);
\draw (F111) -- (F3);
\draw (F2) -- (F21);
\draw (F2) -- (F22);
\draw (F2) -- (F23);
\draw (F2) -- (F24);
\draw (F3) -- (F31);
\draw (F3) -- (F32);
\draw (F3) -- (F33);
\draw (F3) -- (F34);

\draw (F23) -- (F231);
\draw (F23) -- (F232);
\draw (F23) -- (F234);

\draw[dotted, thick] (C) -- (1.3, -2.6);
\draw[dotted, thick] (D) -- (2.3, -2.6);
\draw[dotted, thick] (F22) -- (5.8, -6.25);
\draw[dotted, thick] (F24) -- (7.6, -5.85);
\draw[dotted, thick] (F32) --  (9.5, -6.25);
\draw[dotted, thick] (F33) -- (10.2, -5.85);
\draw[dotted, thick] (F34) -- (11.1, -6.25);

\draw[dotted, thick] (F12) -- (F121);
\draw (F121) -- (F122);
\draw (F121) -- (F123);
\draw[dotted, thick] (F122) -- (11.8, -4.95);
\draw[dotted, thick] (F123) -- (13, -4.95);

\end{tikzpicture}
\caption{Structure of the hash tree for partial verification (schematic representation)}
\label{fig:hash_tree_pv}
\end{figure}

For partial verification, only a portion of the leaves needs to be checked. While the general information from categories 1.--4.\ is required as in the case of complete verification, only some sample of the information for the transition between two checkpoints is verified. This leads us to suggest the following layout for the hash tree: The root hash of the tree is again computed from the concatenation of category-wise root hashes (denoted $h_j$ for category $j$ again). Categories 1.--4.\ should be treated in the same way as for complete verification.\\
For category 5., for each $i_j$ with $j \in \{0, \dots, m\}$, the corresponding data should be hashed together. The top hashes $h_{5,i_j}$ for the different values of $i_j$ should be combined using a binary hash tree whose root hash is the root hash of category 5, $h_5$. A binary tree is suggested, since only some values for $i_j$ are checked and they are not known beforehand. Using a binary tree makes sure that the amount of intermediate information required to recompute $h_5$ is as small as possible. For each $i_j$, we compute the top hash $h_{5, i_j}$ as $h_{5, i_j}=H(i_j|h_{5, i_j, b}|h_{5, i_j, c}|h_{5, i_j, d})$, where $i_j$ is the value from \ subcategory 5.(a), and $h_{5, i_j, b}$, $h_{5, i_j, c}$ and $h_{5, i_j, d}$ are the top hashes of appropriate hash trees combining the data from subcategories (b), (c) and (d), respectively. Since the data from 5.(b), 5.(c) and 5.(d) for any selected $i_j$ need to be checked at the same time, we suggest to use non-binary hash trees in order to save space. The exact number of children at each level of the hash tree can be chosen based on the amount of data to be stored and practical considerations regarding the implementation. \autoref{fig:hash_tree_pv} sketches the proposed hash tree for partial verification.\\

\subsection{Storage requirements}\label{sec:storage}
In this section, we estimate the storage overhead induced by the proposed approach. We assume that the training data themselves are in any case stored by the proving party as a backup or for future use and hence do not consider them in the analysis that follows.\\

Large neural networks can have up to about $2^{27}$ parameters \cite{He2016, Simonyan15}, and storing these parameters can require storage space in the hundreds of megabyte, which we estimate by $2^9\textup{ MB}$ when using 32 bits of precision. Hence, storing only the weights for $m$ checkpoints would require $2^9m \textup{ MB}=\frac{m}{2}\textup{ GB}$ of storage space. The additional storage space for categories 1.--4.\ and 5.(a), 5.(c) and 5.(d) should be negligible in comparison (note that in category 4., only the hash values of the data items $D_i$ are stored, not the data themselves, which might have non-negligible size).\\
For a fixed checkpoint $i_j$, if the hash tree with root hash $h_{5, i_j, b}$ has, for instance, $16$ children at each level, the penultimate level includes $2^{23}$ nodes, each of size $2^5\textup{ B}$ when using SHA-256 as a hash function. Therefore, $2^{28} \textup{ B}=2^8 \textup{ MB}$ of storage space are required to store the penultimate level of the hash tree. The preceding layer would hence require $2^4\textup{ MB}$, and so forth. In this way, the overall storage requirement for these hash trees when using $16$ children per level is less than the one required for storing the parameters themselves. The total storage requirement for storing the data corresponding to $m$ checkpoints is thus bounded by $m\textup{ GB}$.\\
If weight-dependent information for specific optimisation methods needs to be included (e.g.\ for the momentum method \cite{momentum} or Adam \cite{Adam}), this information needs to be stored at every checkpoint. Since it requires essentially the same amount of storage as the weights themselves, the total storage requirement doubles in this case, giving an upper bound of $2m\textup{ GB}$.

\end{document}